\documentclass[showpacs,twocolumn,amssymb,prl,aps]{revtex4}
\usepackage{graphicx,epsfig}
\usepackage{color}

\begin{document}
\title{Influence of nanoconfinement on the nematic behavior of liquid crystals}
\author{Sylwia~Ca{\l}us$^{1}$ }
\author{Daniel~Rau$^2$}
\author{Patrick~Huber$^{2,3}$}
\author{Andriy~V.~Kityk$^{1,3}$ }

\affiliation{$^1$Faculty of Electrical Engineering, Czestochowa
University of Technology, 42-200 Czestochowa, Poland\\
$^2$ FR 7.2 Experimentalphysik, Universit\"at des Saarlandes,
D-66123 Saarbr\"ucken, Germany \\$^3$ Materials Physics and Technology, Hamburg University of Technology, D-21073 Hamburg, Germany}

\date{\today}

\begin{abstract}
We explore the nematic ordering of the rod-like liquid crystals 5CB and 6CB, embedded into parallel-aligned nanochannels in mesoporous silicon and silica membranes as a function of mean channel radius ($4.7 \le R \le 8.3$ nm), and thus geometrical confinement strength, by optical birefringence measurements in the infrared region. The orientational order inside the nanochannels results in an excess birefringence, which is proportional to the nematic order parameter. It evolves continuously upon cooling with a precursor behavior, typical of a paranematic state at high temperatures. These observations are compared with the bulk behavior and analyzed within a phenomenological model. Such approach indicates that the strength of the nematic ordering fields $\sigma$ is beyond a critical threshold $\sigma_c =$ 1/2, that separates discontinuous from continuous paranematic-to-nematic behavior. In agreement with the predictions of the phenomenological approach a linear dependency of $\sigma$ on the inverse channel radius is found and we can infer therefrom the critical channel radii, $R_c$, separating continuous from discontinuous paranematic-to-isotropic behavior, for 5CB (12.1 nm) and 6CB (14.0 nm). Our analysis suggests that the tangential anchoring at the channel walls is of similar strength in mesoporous silicon and mesoporous silica membranes. A comparison with the bulk phase behavior reveals that the nematic order in nanoconfinement is significantly affected by channel wall roughness leading to a reduction of the effective nematic ordering.
\end{abstract}

\maketitle

\section{Introduction}

\begin{figure}[tbp]
\epsfig{file=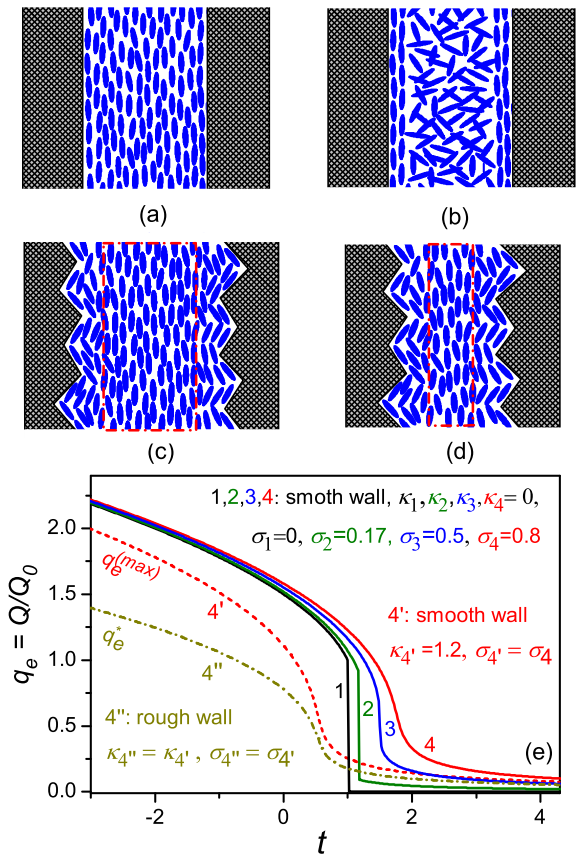, angle=0,
width=0.9\columnwidth}\caption{(color online). Orientational ordering of a nematogen LC inside a cylindric channel. Panels (a) and (b) sketch the nematic ($T<T^*_{IN}$)
 and paranematic ($T>T^*_{IN}$) ordering, respectively, that develop in channels with smooth walls.  Panels (c) and (d) depict the nematic ordering in
broad and narrow channel with rough pore wall, respectively;  { dash-dot rectangles (red online color)} mark the core region of the channel filling, with relatively well developed nematic ordering in comparison to the considerably orientationally disordered near-interface region. Panel (e) shows the order parameter $q_e$ in the KKLZ model
as a function of reduced temperature $t$ for several effective surface fields $\sigma$; traces 1-4,4$'$ correspond to a case of smooth channel wall;
channel wall roughness reduces the effective ordering as sketched by trace 4$''$.} \label{fig1}
\end{figure}

The study of spatial confinement effects on liquid crystals (LCs) embedded in nanoporous and mesoporous materials has been a subject of numerous experimental and theoretical
investigations in the past\cite{Bellini,Iannac,Iannac1,Guegan,Sihai,Kralj,Kutnjak,Kutnjak1,Iannac2,Kutnjak2,Kityk,Kityk1, Steuer2004, Cheung2006, Binder2008}.
It continues to attract considerable interest in many areas of organic electronics \cite{Khoo}, not to mention the widespread use of such hybrid material systems in display technology. Memory effects resulting from frustration, as explored in recent computer simulations \cite{Tanaka} for nematic LCs confined in bicontinuous porous structures, may be considered as one prominent example where orientationally ordered materials are expected to exhibit surprising functionalities because of topological confinement. However, implementation of such ideas in devices based on nanostructured materials requires a better understanding of geometrical confinement and interface coupling that affect spontaneous nematic ordering. This problem has been emphasized in recent molecular dynamic simulations \cite{zanoni1}.

The unusual behavior of confined LCs is often traced to pronounced quenched disorder effects which have been experimentally explored either by embedding of LCs to porous substrates
\cite{Bellini,Iannac,Iannac1,Guegan,Sihai,Kralj,Kutnjak,Kutnjak1} or by dispersion of small particles in LC fluids \cite{Iannac2,Kutnjak2}. Strongly modified phase behavior occurs usually when the system is confined on submicrometer length scales, where the characteristic size of pores or aerosil particles is just a few tens of molecular diameters. In this case the nematic ordering does not develop via a first-order phase transition, typical of the bulk state, but seems to be continuous with an orientational correlation length never increasing beyond the pore size \cite{Bellini}. Consequently, at temperatures even far above the bulk isotropic-to-nematic phase transition there exists a weak residual nematic ordering located preferably in the interface region (see Fig.~1(b)). Therefore it is more appropriate to term the corresponding phase as ''paranematic'' rather than ''isotropic'' \cite{Kralj}. A large number of publications concern the nematogen alkylcyanobiphenyl LC family ($n$CB, $n$=5,7-9,12) embedded in a variety of porous solid substrates, including the random interconnected pore network of controlled pore glass matrices \cite{Kralj,Kutnjak,Kutnjak1} and Vycor glasses \cite{Iannac} or the
parallel-aligned cylindrical pores of aluminum oxide (Anopore) \cite{Iannac1, Grigoriadis2011}, polycarbonate Nuclepore \cite{Crawford} and porous silicon \cite{Guegan}. The results appear to be strongly dependent on the average pore size, types of LC investigated and porous matrix as well as the chemistry (surface-treatment) of the pore walls \cite{Kutnjak1}. Indeed, these studies corroborate that the general behavior of nanoconfined LCs is concomitantly dominated by the anchoring force and the geometric confinement. On a phenomenological level they are described by a single quantity, the so-called \emph{nematic ordering field} $\sigma$. Relevant concepts have been developed by Kutnjak, Kralj, Lahajnar, and Zumer (KKLZ model)\cite{Kutnjak,Kutnjak1}. According to their considerations the nematic ordering field is proportional to the anchoring strength $W$
and inversely proportional to the pore radius $R$.  With increase of the ordering field the nematic ordering becomes more pronounced as it is sketched in Fig.1(e). This effect is shown to be similar to that of an external magnetic field on a spin system which replaces a discrete phase transition by a continuous evolution of the order parameter \cite{Iannac}. Experimentally, such a behavior has been originally demonstrated by NMR measurements on 5CB and 8CB confined to controlled porous glasses of different characteristic void sizes $R$ \cite{Kralj}. However, due to the insufficient accuracy of the experimental data this analysis has been restricted to semi-quantitative evaluations only. Further progress has been achieved in recent high-resolution optical birefringence study on rodlike nematogens 7CB and 8CB\cite{Kityk}  confined in an array of parallel-aligned nanochannels of monolithic silica membranes. These matrices turn out to be particularly suitable for a very precise determination of the order parameter behavior. However, the measurements in Ref. \cite{Kityk} have been performed for one channel radius only ($2R$=10 nm), thus a challenging issue remains a verification of the KKLZ model, particularly the $R$-dependence of the ordering field $\sigma$ in the nanoscale region.

In this paper we explore the nematic ordering of the model nematogen LCs 5CB (4-Cyano-4'-pentylbiphenyl) and 6CB (4-cyano-4'-hexylbiphenyl) embedded in parallel-aligned nanochannels of size-controlled porous silicon ($p$Si) and porous silica ($p$SiO$_2$) membranes studied by optical birefringence measurements. The experimental results obtained are compared with the bulk behavior of these LCs and analyzed within the KKLZ approach. { The choice of these low-molecular LC compounds is motivated for the two following reasons. First, in the bulk state both thermotropic liquids are characterized by quite simple phase diagrams. Upon cooling they exhibit a single phase transition before solidification,  i.e. a transition from the isotropic (I) to the nematic (N) phase at $T_{IN}\approx$ 308.5 K (5CB) and $T_{IN}\approx$302.4 K (6CB) \cite{Dn,Janik}, respectively. Second, the strength of the first-order character of the isotropic-to-nematic transition practically is not affected by the confinement as has been ascertained in recent theoretical studies \cite{Kralj10,Kralj11}. The latter essentially simplifies the analysis based on the KKLZ model.}

\section{Experimental}

Nematogen LCs 6CB and 5CB have been purchased from Merck.  The $p$Si membranes have been prepared by electrochemical anodic etching of highly p-doped,  $\langle$100$\rangle$ oriented silicon wafers. To obtain membranes of $p$SiO$_2$, the free standing $p$Si membranes have been subjected to further thermal oxidation for 12 h at $T$=800 $^o$C under standard atmosphere. By applying different etching conditions we have obtained  membranes with average pore radius 4.7$\pm$0.3 nm ($p$Si, porosity $P$=60\%), 6.3$\pm$0.3 nm ($p$SiO$_2$, $P$=50\%) and  8.3$\pm$0.3 nm ($p$Si, $P$=67\%) as determined by recording of volumetric N$_2$-sorption isotherms at $T$=77 K. The mesoporous membranes were completely filled by capillary action (spontaneous imbibition) of LCs in the isotropic phase  \cite{Gruener2010}. For the bulk measurements the sample cells, made of parallel glass plates ($d$=30 $\mu$m), have been filled by the LCs providing homogeneous (5CB) or homeotropical (6CB) alignments.

Note that pSi-membranes of about 300 $\mu$m thickness, as used in our studies, are not suitable for optical polarization measurements in the visible spectral region due to a strong fundamental light absorption at the wavelengths below $\sim$1000~nm. For this reason the optical birefringence measurements were performed at $\lambda$ =1342 nm in the infrared region, for which both pSi and pSiO$_2$ membranes, the glass sample cell employed for the bulk measurements and both host LC compounds, are optically well transparent.

The samples were tilted out with respect to their long channel axis (optical axis) by an angle of 40-45 deg. The optical polarization setup employs a photoelastic modulator and a dual lock-in detection system, as described in Ref.\cite{Kityk}, providing an accuracy of the optical retardation measurements better than 5$\cdot$10$^{-3}$ deg.

\begin{figure*}[tbp]
\epsfig{file=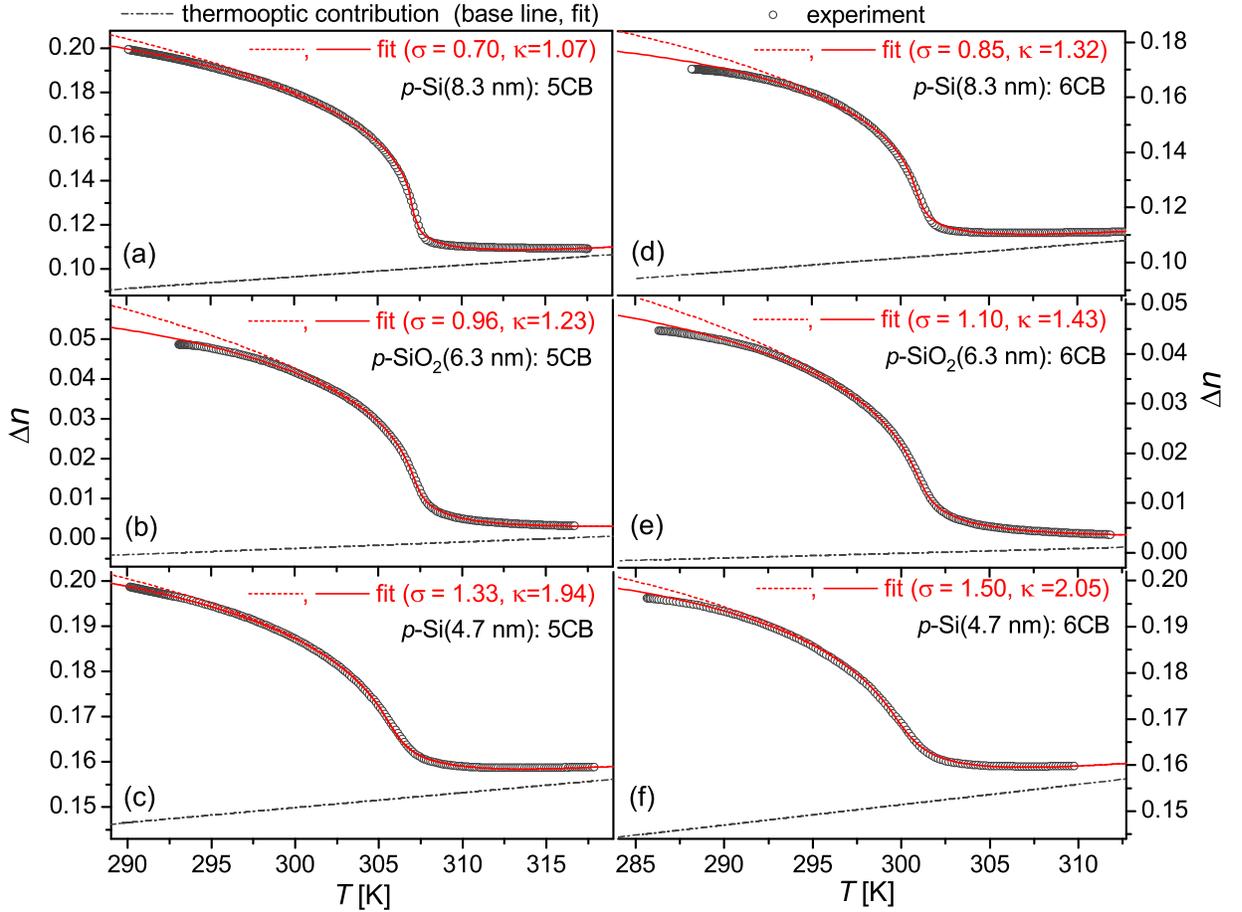, angle=0,width=1.9\columnwidth}
 \caption{(color online). Measured optical birefringence $\Delta n$ ($\lambda=$1342 nm) of nematogen 5CB (left panels) and 6CB (right panels) embedded in parallel-aligned nanochannels
 of several porous membranes: (a, d) $p$Si ($P$=68\%, $R=8.3$ nm); (b, e) $p$SiO$_2$ ($P$=50\%, $R=6.3$ nm); (c,f) $p$Si ($P$=60\%, $R=4.7$ nm). { The broken line (red online color)} represents the best fit obtained within the KKLZ-model based on a free energy expansion (\ref{eq1}).
Completion of the free energy by higher order terms (up to tenth order) describes the saturation of the order parameter at lower temperatures more accurately, see { solid line (red online color)}.  The base line corresponding to the thermooptic contribution {(dash-dot line, grey online color)} is obtained within the same fitting procedure. The values of the effective nematic ordering field $\sigma$ and quenched disorder factor $\kappa$ are quoted for each LC/porous matrix combination.}
 \label{fig2}
\end{figure*}

\section{Results and discussion}

In Fig.~2 we show the temperature dependences of the optical birefringence measured for the mesoporous membranes with different pore radii filled by LCs 5CB and 6CB, respectively.
The nematic ordering does not occur via discontinuous, first-order phase transitions, as observed in the bulk state (see Fig.~3), but evidently evolves continuously in the vicinity of the temperature corresponding to the isotropic-to-nematic bulk transition $T_{IN}$.

Consequently, at temperatures above $T_{IN}$ there is a specific precursor behavior indicating a residual nematic ordering at high temperature, typical of a paranematic state. The continuous character of the paranematic-to-isotropic transition can be described by a sufficiently large nematic ordering field $\sigma>\sigma_c=0.5$ (due to the small channel diameters), if one refers to the KKLZ approach. By the same token a comparison of our measurements with the predictions of this model allows us to determine $\sigma$ for each channel radius $R$ as outlined below.

The untreated surface of silica and silicon pores enforces planar anchoring without a preferred lateral direction (within the plane) \cite{Drevensek2003, zanoni1}, whereas the elongated geometry of cylindrical pores yields additionally a preferred orientation of molecules along the pores axes. In accordance to the KKLZ model it is convenient to characterize the relevant ordering by the scaled order parameter $q=Q/Q_0$, where $Q=\frac{1}{2}\langle 3 \cos^2\theta-1\rangle$ stands for the nematic (orientational) order parameter in its classical definition, $Q_0=Q(T_{IN})$ and $\theta$ is the angle between the long molecular axis and the director. The dimensionless free energy density of a nematic LC in confined geometry reads then:
\begin{equation}
f=tq^2-2q^3+q^4-q\sigma+\kappa q^2 \label{eq1}
\end{equation}
{ where $t=(T-T^*)/(T_{IN}-T^*)$ is the dimensionless reduced temperature, $T^*$ is the effective temperature} and the $\kappa$-term completes the free energy expansion \cite{Kutnjak,Kutnjak1} by a contribution originating from quenched disorder effects. Minimization of Eq.(\ref{eq1}) with respect to $q$ gives its equilibrium magnitude $q_e$ as it is sketched in Fig.\ref{fig1}(e). The isotropic (paranematic)-to-nematic transition occurs at $t_n=1+\sigma-\kappa$ and is discontinuous as long as $\sigma \le \sigma_c=0.5$. Above this value a continuous transition is realized.

The orientational order inside the pores results in an excess birefringence, $\delta(\Delta n) \propto q_e$, which appears on the background of the geometric birefringence, $\Delta n_g$, being a characteristic feature of membranes with parallel nanochannels \cite{Kityk2}. The measured birefringence represents a superposition of these two contributions, i.e. $\Delta n=\delta(\Delta n)+\Delta n_g$. The geometric birefringence is quite small { for the entirely filled $p$SiO$_2$-membrane due to similar refractive indices of host silica ($n\approx$1.46) and isotropic (or average) refractive indices of the guest 5CB or 6CB LCs ($n\approx$1.58 \cite{Dn}). For this reason it has been ignored in earlier studies \cite{Kityk}.
However, the geometric birefringence is considerable in the case of entirely filled  $p$Si membranes because of the large refractive index of the silicon host ($n\approx$3.4 \cite{Pap}).  Thus $\delta(\Delta n)$ and $\Delta n_g$ appear to be comparable,} as it is evident from Fig.~\ref{fig2}(c,f). Moreover, $\Delta n_g$ exhibits a smooth temperature behavior
(referred to as "thermooptic" changes in the following) due to the varying refractive indices of the host Si (SiO$_2$) matrix and the guest LC as well as changes of the porosity caused by thermal expansion/contraction of the porous matrix. Accordingly, this "thermooptic" contribution, approximated by a linear dependence, has been included in the fitting procedure, see grey dash-dot line in Fig.~\ref{fig2}. { The fitting procedure is based on an iterative, numerical minimization of Eq.(1)  to extract $\sigma$ and $\kappa$ values that provide smallest least-square deviations between the calculated curve and the experimental points. Similarly to Ref.\cite{Kutnjak} we first calibrated the temperature scale of the KKLZ model by analyzing the bulk birefringence in the region of the isotropic-to-nematic transition. This analysis gives $T_{IN}-T^*$ approximately equal 3.4 K (5CB) and 3.2 K (6CB), resp., i.e. values close to the magnitude of 3.5 K  obtained for 8CB \cite{Kutnjak}. In fact the difference $T_{IN}-T^*$, which defines the proximity of the system to the tricritical point, may vary in general under confinement conditions, as has been recently emphasized by Kralj \emph{et al}\cite{Kralj10,Kralj11}. This effect appears to be dominant if surface wetting interactions are weak, which is observed in samples containing relatively highly flexible LCs molecules, as e.g. 12CB. However, it is expected to be weak in 5CB and 6CB, where surface wetting strengths play important roles. This particularly justifies a choice of these LCs  for the present experiments. Because of this reason the temperature conversion factor, defined by the difference $T_{IN}-T^*$, has been fixed in all consequent fitting procedures to values specified above.} Dashed red curves in these figures represent the best fits of measured $\Delta n(T)$-dependences based on the free energy expansion (\ref{eq1}). Including higher order terms (up to the tenth order) into Eq.(\ref{eq1}) better reproduces the saturation of the nematic order parameter at low temperatures (solid red lines in Fig.~\ref{fig2}), but has no influence { on the fit quality in the transition region or the extracted $\sigma$ and $\kappa$ values. Insignificant deviations between the experimental points and fitting curves may be recognized also directly in the transition region. In most cases the fits exhibit here slightly steeper changes than the experimental data. The reason for such discrepancies could originate in pore wall roughness as well as the pore diameter distribution, which are both not considered in the KKLZ model and which should lead to a more smeared behavior in the transition region.}

\begin{figure}[tbp]
\begin{flushleft}
  \epsfig{file=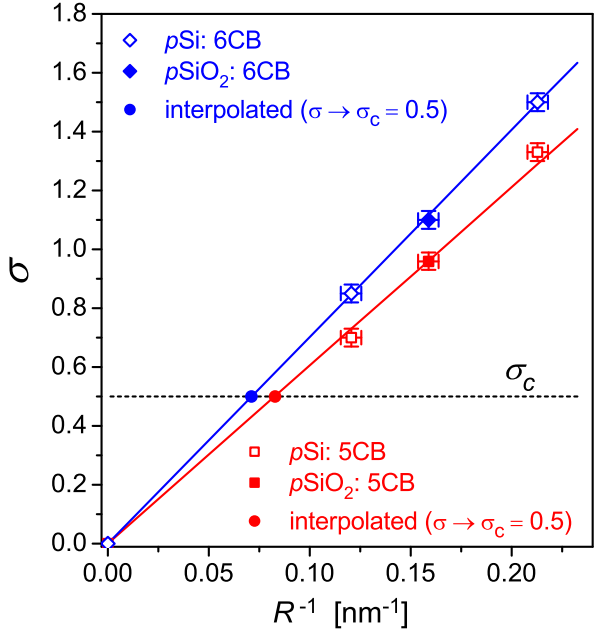, angle=0, width=0.95\columnwidth}
  \caption{(color online). Nematic ordering field $\sigma$ vs inverse mean channel radius $R^{-1}$. Square symbols correspond to $\sigma$ values as extracted within the fitting analysis of the $\Delta n(T)$-dependences
  shown in Fig.~\ref{fig2}. Solid lines are the best linear fits of the $\sigma$ data points. Their intersections with a horizontal line, $\sigma=\sigma_c=1/2$
  (see solid circles) yield coordinates of the critical points $\sigma_c(R_c=12.1$ nm$)$ (5CB) and $\sigma_c(R_c=14.0$ nm$)$ (6CB) as discussed in the text. } \label{fig4}
\end{flushleft}
\end{figure}

In Fig.~\ref{fig4} we show $\sigma$ vs $R^{-1}$. It is for both confined LCs investigated practically linear. Following the considerations in Refs. \cite{Kutnjak,Kutnjak1} the effective nematic ordering field $\sigma$ is expressed as:
\begin{equation}
\sigma=\frac{2\xi^{*2}W_{n1}}{RkQ_0} \label{eq2}
\end{equation}
where $\xi^*=\xi(T_{IN})$ is the correlation length at the temperature of the isotropic-to-nematic transition, $W_{n1}$ is the anchoring strength and $k$ is the nematic elastic constant. Note that we neglect here the first term of Eq. (7c) of Ref.\cite{Kutnjak1} which is $\propto R^{-2}$. In the case of tangential ordering it is expected to be small. For this reason the $\sigma(R^{-1})$-dependence in Fig.~\ref{fig4} is dominated for each LC  by the linear contribution in accordance with Eq.(\ref{eq2}). Interestingly, for both LCs the data point $\sigma(R=$6.3 nm$)$ determined for the $p$SiO$_2$ membrane lies perfectly on the linear fit which passes through two other data points, $\sigma(R=$4.7 nm$)$ and $\sigma(R=$8.3 nm$)$ that correspond to $p$Si membranes. This implies an identical anchoring strength for the silicon and silica membrane, which appears surprising at first glance. We suggest that the native oxide (SiO$_2$) layer usually forming on silicon surfaces renders the inner surfaces of the silicon membrane silica-like and thus the fluid/wall interaction, which determines the anchoring strength, is identical in porous silicon and silica. Slightly different slopes of the $\sigma(R^{-1})$-dependences, as one can see in Fig.~\ref{fig4} for
nanoconfined 5CB and 6CB LCs, originate from the difference in material constants characterizing these LCs and, likewise, the difference in anchoring forces. The slope of the $\sigma(R^{-1})$-dependence is defined by the factor $\varrho=2\xi^{*2}W_{n1}/(kQ_0)$. The fit of the data points in Fig.~\ref{fig4} yields $\varrho$-factor of 6.05 nm (5CB) and 7.03 nm (6CB), respectively. With the typical correlation length $\xi^*\sim$10-15 nm, elastic constant $k=$7.7$\cdot$10$^{-12}$ N (5CB),
$k=$8.0$\cdot$10$^{-12}$ N (6CB) \cite{elastic} and $Q_0\sim$1/2 one yields an anchoring strength $W_{n1}$ in the range 50-100 $\mu$J/m$^2$. These values are of the same order of magnitude as obtained for 8CB in random pore networks of controlled pore glass \cite{Kutnjak1},  i.e. the interaction of 5CB or 6CB molecules with SiO$_2$ surface corresponds to a regime of weak anchoring.

\begin{figure}[tbp]
\begin{flushleft}
\epsfig{file=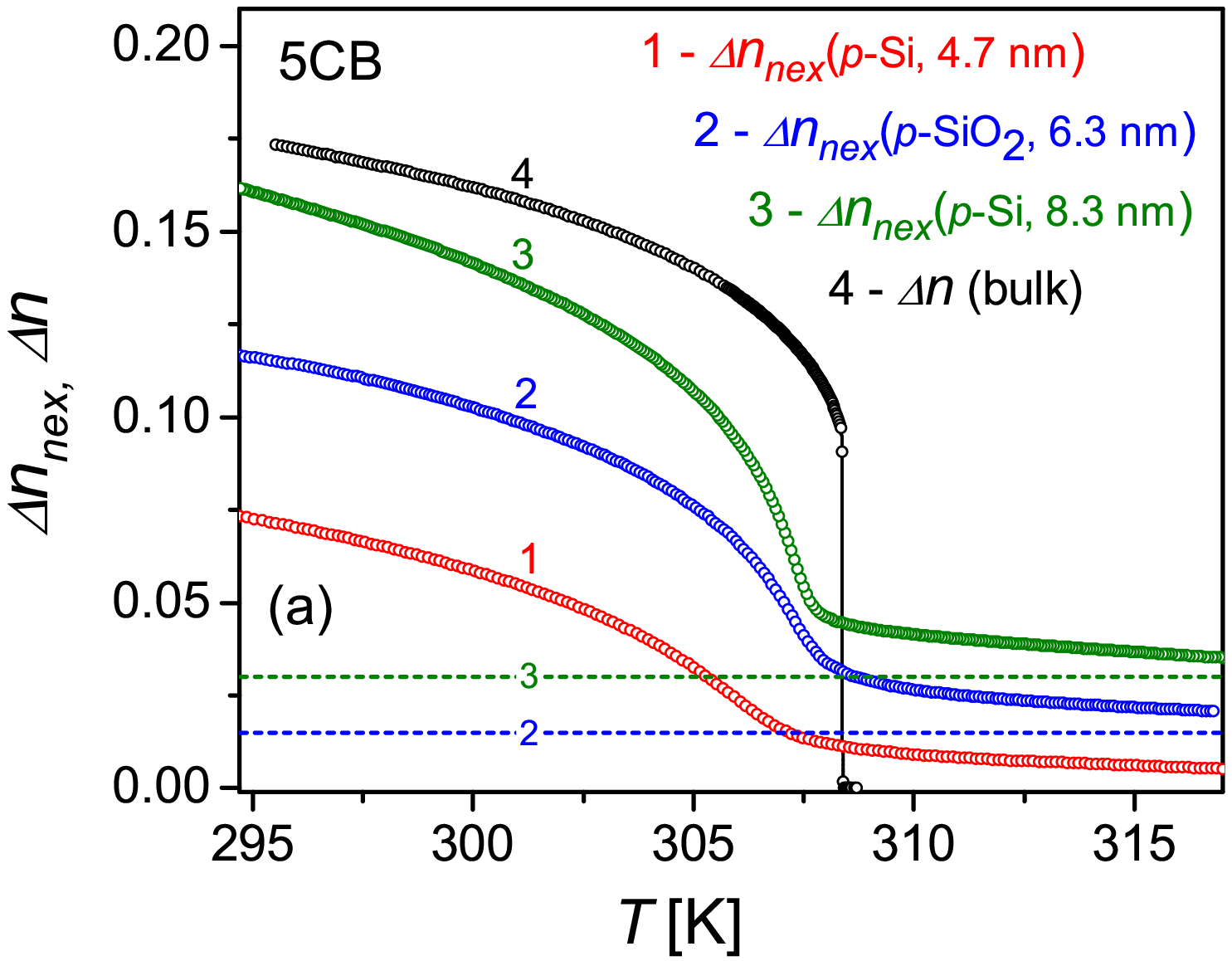, angle=0, width=0.95\columnwidth}
  \epsfig{file=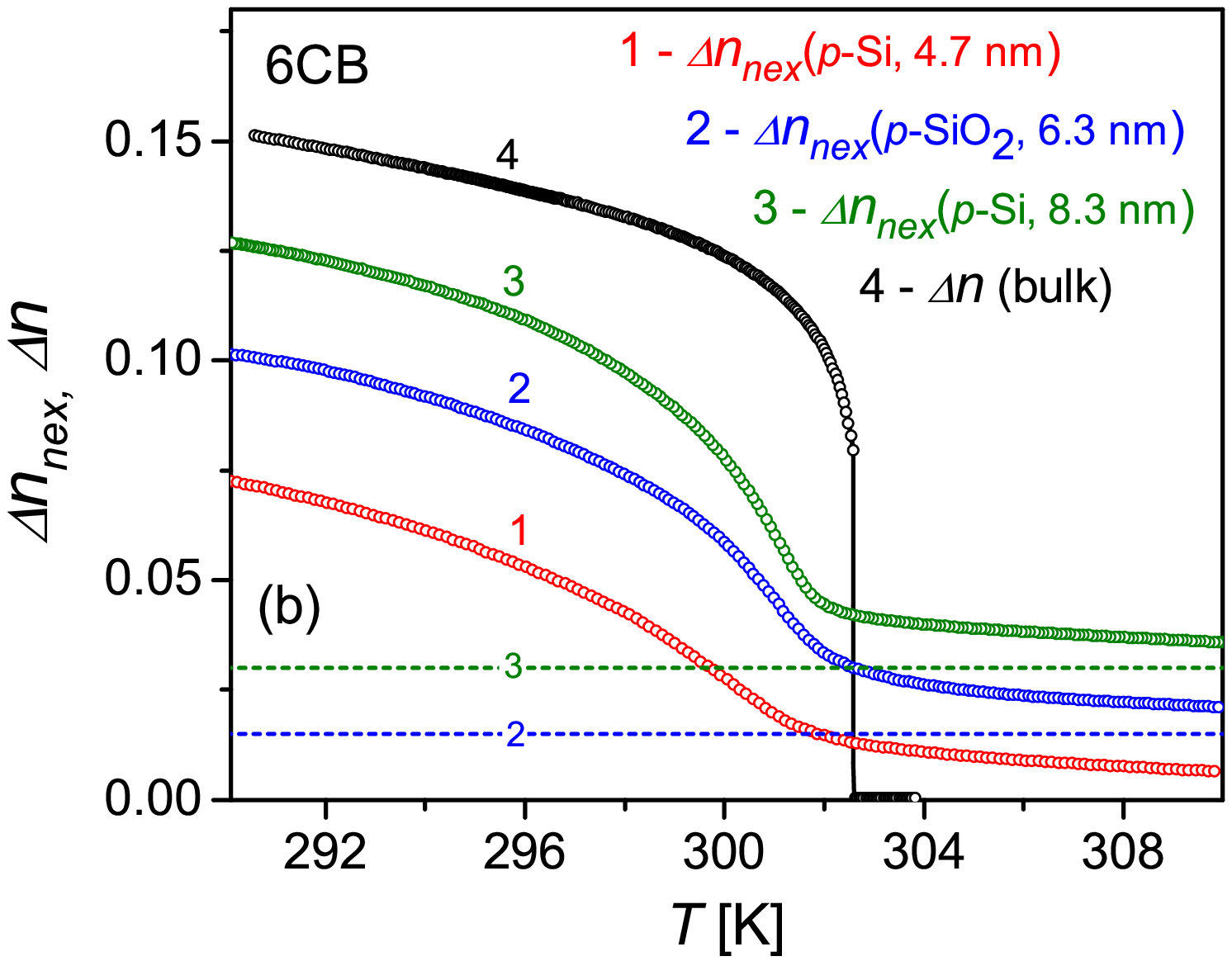, angle=0,width=0.97\columnwidth}
\caption{(color online). Normalized excess birefringence $\Delta n_{nex}$ vs temperature $T$ (blue, green and red colors) of
nematogen 5CB [panel (a)] and 6CB [panel (b)] embedded into mesoporous membranes, as calculated from the data presented in
Fig.~\ref{fig2}. The traces $\Delta n_{nex}(T)$ for $p$SiO$_2$ (6.3 nm, blue color) and $p$Si (8.3 nm, green color)
 are shifted vertically by 0.015 and 0.03, respectively, in order to avoid their overlap; horizontal broken lines { with relevant labels} marks 0-levels for these
traces. Bulk birefringence (black circles) is given for comparison.} \label{fig5}
\end{flushleft}
\end{figure}

\begin{figure}[tbp]
\begin{flushleft}
  \epsfig{file=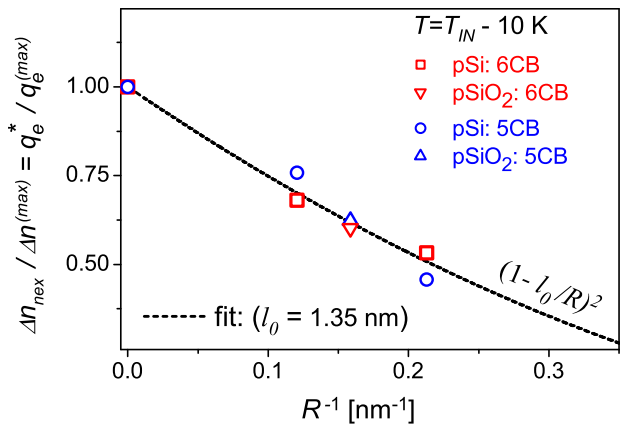, angle=0, width=0.95\columnwidth}
  \caption{(color online). $\Delta n_{nex}/\Delta n^{(max)}$ vs $R^{-1}$ as determined for both LCs at
  $T=T^*_{IN}-10$ K using the normalized excess birefringence, $\Delta n_{nex}(T)$, and bulk birefringence, $\Delta n(T)$, data in Fig.~\ref{fig5}.} \label{fig6}
\end{flushleft}
\end{figure}

Interpolation of the $\sigma (R^{-1})$ dependences to $\sigma_c$=0.5 (see Fig.\ref{fig4})  allows one to determine the critical pore radii $R_c$  12.1 nm  for 5CB and 14.0 nm for 6CB, respectively, see solid circles at the intersection of relevant lines. These values are close to the ones obtained for other $n$CB LCs, e.g. 7CB ($R_c$=8 nm)  or 8CB ($R_c$=11.5 nm) embedded into $p$SiO$_2$ membranes \cite{Kityk}.

By subtracting the thermooptic contribution, $\Delta n_g(T)$, from the measured birefringence $\Delta n(T)$ shown in Fig.~\ref{fig2} one obtains the excess birefringence, $\delta (\Delta n)\propto q_e$, originating exclusively from the orientational ordering of the confined molecules. The excess quantity $\delta (\Delta n)$ roughly scales with the porosity of the membrane, $P$, thus a comparison of the orientational ordering in samples of different porosity or with the bulk state requires its normalization by $P$. In a more accurate characterization, which is based on the Bruggeman
approximation, the normalized excess birefringence is given by $\Delta n_{nex}=\zeta\delta (\Delta n)/P$, where $\zeta$  is a correction factor that depends on composite parameters, i.e. the actual pore geometry, the porosity and the difference between the refractive indices of the host material and the guest LC. If the latter is small, as e.g. for the composites $p$SiO$_2$+$n$CB ($n$=5-10), the factor $\zeta$ deviates only very slightly from 1.0 ($\zeta=$0.993-0.994) and the corresponding correction can be ignored. This, however, is not the case for the composites with considerably different refractive indices of the host and guest components as, e.g., $p$Si+$n$CB for which $\zeta$ equals 0.943 ($P$=60\%) and 0.909 ($P$=68\%) as provided by estimations based on the
Bruggeman approach for membranes with strongly elongated parallel channels; for details regarding the methodology of such evaluations see Ref.\cite{Kityk2}. The normalized excess birefringence $\Delta n_{nex}$ of nanoconfined LCs, determined in such a way, is shown in Fig.~\ref{fig5} and is compared with the bulk behavior (black color).

In general, a decreasing channel radius leads to an increase of the effective nematic ordering field $\sigma$. Accordingly, enhancing of the nematic ordering is expected in the entire
temperature region as shown in Fig.~\ref{fig1}(e);  see the sequence of solid green, blue and red lines corresponding to a rising sigma value ($\kappa$=0; $\sigma_2 < \sigma_3 <  \sigma_4$, respectively). On the other hand, quenched disorder effects, represented by the $\kappa$-term in the free energy (\ref{eq1}) lead to a lowering of the paranematic-to-nematic  transition as it is illustrated by the dashed red line in Fig.~\ref{fig1}. On the reduced temperature scale it leads just to a temperature shift of the whole curve by $\Delta t=-\kappa$. Thus depending on the relative magnitude of $\kappa$ and $\sigma$ the orientational ordering at a certain fixed temperature below $T_{IN}$ may even decrease while the effective nematic ordering field rises.

Distortions caused by surface roughness of the pore walls, as sketched in Fig.~\ref{fig1}(c)(d), lead to further decrease in the effective (averaged)  nematic ordering. Accordingly, the dashed curve  4$'$ in Fig.~\ref{fig1}(e) in fact corresponds to the theoretically achievable maximum value $q_e^{(max)}$ for, so to say, ideally smooth channel walls [Fig.~\ref{fig1}(b)], whereas the dash-dot curve 4$''$ accounts for disordering due to channel wall roughness [Fig.~\ref{fig1}(c),(d)] leading thus to its reduced magnitude $q_e^{*}$. By $\Delta n^{(max)}$ we denote the birefringence relevant to $q_e^{(max)}$; it can be evaluated for each pore geometry and temperature using the measured bulk birefringence (Fig.~\ref{fig5} (a)(b)) and the extracted fit parameters, $\sigma$ and $\kappa$. On the other hand,  $\Delta n_{nex}$ is relevant to  $q_e^{*}$. In such a consideration the ratio $q_e^{*}/q_e^{(max)}$ characterizes the degree of confined nematic order and is equal to the ratio $\Delta n_{nex}/\Delta n^{(max)}$. In Fig.~\ref{fig6} we present $\Delta n_{nex}/\Delta n^{(max)}$ vs $R^{-1}$ as determined for both LCs at $T=T_{IN}-10$ K. The effective orientational order evidently rises with increasing  pore diameter. Such a trend can be rationalized by a consideration of the contributions that originate from different regions of a pore filling where relatively well developed nematic order in the core region competes with orientational disorder of few molecular layers next to the pore walls, as sketched in Fig.~\ref{fig1},(c) and (d). Assuming that the nematic order is fully developed in the core region and is negligible in the near-interface region the degree of nematic order inside the pore $q_e^{*}/q_e^{(max)} = (1-l_0/R)^2$, where $l_0$ is the effective thickness
of the disordered near-interface layer. Here $l_0$ is of the order of few molecular layers satisfying the inequality, $l_0 \le R_0$. For large  pore radii the effective nematic order is favored by the ordered core component ($q_e^{*}/q_e^{(max)}|_{R\rightarrow\infty}$ = 1 ),  whereas at small ones the disordered near-interface layers provide a dominant contribution ($q_e^{*}/q_e^{(max)}|_{R\rightarrow l_0}$ = 0 ). The dashed black line in Fig.~\ref{fig6} represents the best fit of the experimental data points by the $(1-l_0/R)^2$-function which yields an effective thickness of the disordered near-interface layer of $l_0=$1.35 nm, corresponding to approx.  2 to 3 intermolecular spacings $a_0$ (for 5CB $a_0=5.0$ {\AA} \cite{Komolkin}). One should emphasize that $l_0$ represents here the effective model parameter rather than the thickness of the real near-interface layer. In reality the near-interface region exhibits a certain nematic ordering. Thus, presumably, our case corresponds to much weaker distortions of the nematic order that start in the layer next to the pore wall and radially decays on distances comparable with the orientational correlation length, i.e. at least over distances several times longer than $l_0$. Such a radial gradient in the orientational order has been inferred in Monte Carlo simulations on rod-like liquid crystals confined in nanochannels \cite{Ji2009, Ji2009a}.

\section{4. Conclusion}
We have reported an optical birefringence study with infrared Laser light on the nematic ordering of the nematogen LCs 5CB and 6CB, embedded in parallel-aligned nanochannels of channel-radius-controlled porous silicon and silica membranes ($4.7 \le R \le 8.3$ nm). These results are compared with the bulk behavior and analyzed within a phenomenological approach.

The molecular orientational order inside the channels gives rise to an excess birefringence, which is proportional to the nematic order parameter and appears on top of the background of the geometric birefringence typical of a homogeneous matrix permeated by parallel-aligned nano channels. By applying an appropriate fitting procedure we were able to separate these distinct birefringence contributions and thus to determine the magnitude of the effective nematic ordering field $\sigma$ originating from the geometric confinement.

The nematic ordering in the cylindrical nanochannels evolves continuously with characteristic precursor behavior typical of the paranematic state. The strength of the ordering fields, imposed by the silicon or silica walls, appears in all cases investigated beyond a critical threshold $\sigma_c =$ 1/2 , that separates discontinuous from continuous paranematic-to-nematic behavior. An interpolation of the linear $\sigma (R^{-1})$-dependence provides values for the critical pore radius, $R_c = 12.1$ nm (5CB) and $R_c=14.0$ nm (6CB). Those values are close to the ones obtained recently for 7CB and 8CB embedded int porous silica membranes.

Our analysis evidently suggests that the tangential anchoring strength acting at the channel walls is the same in porous silicon and silica membranes, presumably because of the formation of a native oxide layer at the inner surfaces of the porous silicon membrane. A comparison with the bulk behavior suggests that the nematic order of 5CB and 6CB in nanoconfinement is significantly affected by the roughness of the channel walls. The impact appears to be more pronounced for narrow pores ($R <$ 5 nm), where the reduction in nematic ordering exceeds even 50\%.

\section{Acknowledgement}
The authors acknowledge financial support by the joint French-German, ANR-DFG program "TEMPLDISCO", DFG Grant No. Hu850/3-1.


\begin{thebibliography}{00}

\bibitem{Bellini} T.~Bellini, N.~A.~Clark, C.~D.~Muzny, L.~Wu, C.~W.~Garland,
D.~W.~Schaefer, B.~J.~Olivier, Phys. Rev. Lett. \textbf{69}, 788 (1992).

\bibitem{Iannac} G.~S.~Iannacchione, G.~P.~Crawford, S.~Zumer, J.~W.~Doane, D.~Finotello,
 Phys. Rev. Lett. \textbf{71}, 2595 (1993).

\bibitem{Iannac1} G.~S.~Iannacchione, J.~T.~Mang, S.~Kumar, D.~Finotello,
 Phys. Rev. Lett. \textbf{73}, 2708 (1994).

\bibitem{Guegan} R.~Guegan, D.~Morineau, R.~Lefort, A.~Moreac, W.~Beziel, M.~Guendoz,
J.~M.~Zanotti, B.~Fric, J. Chem. Phys. \textbf{126}, 064902 (2007).

\bibitem{Sihai} S.~Qian, G.~S.~Iannacchione, D.~Finotello, Phys. Rev. E \textbf{57}, 4305 (1998).

\bibitem{Kralj} S.~Kralj, A.~Zidansek, G.~Lahajnar, S.~Zumer, R.~Blinc, Phys. Rev. E \textbf{57}, 3021 (1998).

\bibitem{Kutnjak} Z.~Kutnjak, S.~Kralj, G.~Lahajnar and S.~Zumer, Phys. Rev. E \textbf{68}, 021705 (2003).

\bibitem{Kutnjak1} Z.~Kutnjak, S.~Kralj, G.~Lahajnar and S.~Zumer, Phys. Rev. E \textbf{70}, 051703 (2004).

\bibitem{Iannac2} G.~S.~Iannacchione, C.~W.~Garland, J.~T.~Mang, T.~P.~Rieker, Phys. Rev. E \textbf{58}, 5966 (1998).

\bibitem{Kutnjak2} Z.~Kutnjak, S.~Kralj, and S.~Zumer, Phys. Rev. E \textbf{66}, 041702 (2002).

\bibitem{Kityk} A.~V.~Kityk, M.~Wolff, K.~Knorr, D.~Morineau, R.~Lefort, P.~Huber, Phys. Rev. Lett. \textbf{101}, 187801 (2008).

\bibitem{Kityk1} A.~V.~Kityk and P.~Huber, Appl. Phys. Lett. \textbf{97}, 153124 (2010).

\bibitem{Steuer2004} H. Steuer, S. Hess, and M. Schoen, Phys. Rev. E \textbf{69}, 031708 (2004).

\bibitem{Cheung2006} D. Cheung and F. Schmid, Chem. Phys. Lett. \textbf{418}, 392 (2006).

\bibitem{Binder2008} K. Binder, J. Horbach, R. Vink, A. De Virgiliis, Soft Matter  \textbf{4}, 1555 (2008).

\bibitem{Khoo} I.~C.~Khoo, Phys. Rep.-Rev. Sec. Phys. Lett. \textbf{471}, 221 (2009).

\bibitem{Tanaka} T.~Araki, M.~Buscaglia, T.~Bellini, H.~Tanaka, Nature Materials, \textbf{10}, 303 (2011).

\bibitem{zanoni1} A.~Pizzirusso, R.~Berardi, L.~Muccioli, M.~Ricci, C.~Zannoni, Chem. Sci., \textbf{3}, 573 (2012).

\bibitem{Grigoriadis2011} C. Grigoriadis, H. Duran, M. Steinhart, M. Kappl, H.J. Butt, G. Floudas, ACS Nano \textbf{5}, 9208 (2011).

\bibitem{Crawford} R.~J.~Ondris-Crawford, G.~P.~Crawford, J.~W.~Doane, S.~Zumer, M.~Vilfan, I.~Vilfan, Phys. Rev. E  \textbf{48}, 1998 (1993).

\bibitem{Dn} I.~Chirtoc, M.~Chirtoc, C.~Glorieux, J.~Thoen, Liq. Crystals, \textbf{31}, 229 (2004).

\bibitem{Janik} J.~Janik, R.~Tadmor, J.~Klein, Langmuir \textbf{13}, 4466 (1997).

{ \bibitem{Kralj10} S.~Kralj, R.~Rossob, E.~G.~Virga, Soft Matter, \textbf{7}, 670 (2011).

\bibitem{Kralj11} S.~Kralj, G.~Cordoyiannis, D.~Jesenek, A.~Zidan{\v{s}}ek, G.~Lahajnar, N.~Novak, H.~Amenitsch, Z.~Kutnjak, Soft Matter, \textbf{8}, 2460 (2012).}

\bibitem{Gruener2010} S.~Gruener and P.~Huber, J. Phys.: Cond. Matt. \textbf{23}, 184109 (2010).

\bibitem{Drevensek2003} I. Drevensek Olenik, K. Kocevar, I. Musevic, and Th. Rasing, Eur. Phys. J. E \textbf{11}, 169 (2003).

\bibitem{Kityk2} A.~V.~Kityk, K.~Knorr, P.~Huber, Phys. Rev. B, \textbf{80}, 035421 (2009).

{ \bibitem{Pap} A.~E.~Pap, K.~Kord{\'a}s , J.~V{\"a}h{\"a}kangas, A.~Uusim{\"a}ki, S.~Lepp{\"a}vuori, L.~Pilon, S.~Szatm{\'a}ri, Opt. Mater. \textbf{28}, 506 (2006).}

\bibitem{elastic} H.~Hakemi, E.~F.~Jagodzinski, D.~B.~DuPr{\'e}, J. Chem. Phys. \textbf{78}, 1513 (1983).

\bibitem{Komolkin} A.~V.~Komolkin, A.~Laaksonen, A.~Maliniak, J. Chem. Phys. \textbf{101}, 4103 (1994).

\bibitem{Ji2009} Q.~Ji, R. Lefort, R. Busselez, D. Morineau,  J. Chem. Phys. \textbf{130}, 234501 (2009).

\bibitem{Ji2009a} Q.~Ji, R. Lefort, D. Morineau,  Chem. Phys. Lett. \textbf{478}, 161 (2009).

\end{thebibliography}
\end{document}